# R-Shiny Applications for Local Clustering to be Included in the `growclusters` for R Package

Randall Powers*, Terrance Savitsky*, and Wendy Martinez+

*Office of Survey Methods Research, U.S. Bureau of Labor Statistics
+U.S. Census Bureau

Powers.Randall@bls.gov

**Abstract:** `growclusters` for R is a package that estimates a partition structure for multivariate data. It does this by implementing a hierarchical version of *k*-means clustering that accounts for possible known dependencies in a collection of datasets, where each set draws its cluster means from a single, global partition. Each component data set in the collection corresponds to a known group in the data. This paper focuses on R Shiny applications that implement the clustering methodology and simulate data sets with known group structures. These Shiny applications implement novel ways of visualizing the results of the clustering. These visualizations include scatterplots of individual data sets in the context of the entire collection and cluster distributions versus component (or sub-domain) datasets. Data obtained from a collection of 2000-2013 articles from the Bureau of Labor Statistics (BLS) Monthly Labor Review (MLR) will be used to illustrate the R-Shiny applications. Here, the known grouping in the collection is the year of publication.

**Disclaimer**: Any opinions and conclusions expressed herein are those of the authors and do not reflect the views of the Bureau of Labor Statistics or the U.S. Census Bureau.

**Key words:** clustering; R Shiny; unsupervised learning; *k*-means clustering, text analysis

## 1. Background

Clustering is the process of grouping data such that records or observations assigned to the same cluster are more similar as compared to data points or observations in other groups. This process is a key aspect of exploratory data analysis. Furthermore, clustering is an iterative process that can use a variety of algorithms and approaches [Martinez, et al., 2011]. So, the data scientist should try different clustering methods to search for structure or groups in the data. [Powers, et al., 2022]

Several R packages containing functions for clustering exist on CRAN. The `growclusters` package (under development and coming out soon) is another package that implements a novel clustering methodology based on hierarchical Bayesian models. It is designed to estimate a partition structure for relatively high-dimensional multivariate data [Savitsky, 2016].

Given the iterative nature of exploratory data analysis, the creation of an interactive data visualization tool to accompany the `growclusters` package was a top priority and inspiration for this project. Building R Shiny applications was determined to be the best solution to achieve this goal.

This project is ongoing; currently we are finishing work on three R Shiny applications to accompany the package. The first, called `gendata`, allows the user to create a customized input data set to be used in the second R Shiny app. This app, called `dpGrowclusters` will perform the clustering. The third R Shiny app implements the hierarchical version of `growclusters` and is called `hdpGrowclusters`. This paper will summarize the functionality and usefulness of each app.

A corpus of Bureau of Labor Statistics (BLS) *Monthly Labor Review* (MLR) articles from 2000 to 2013 [Powers, et al., 2021] will be used as the source material to construct a dataset to demonstrate the functionality of the clustering apps. The data are not the focus of this article; they are only used to illustrate the applications. See Powers, et al. [2021] for details on the data and how the dataset was constructed. For the application in this paper, each MLR article was converted to a computable form, and each data point or observation referred to in the examples corresponds to an MLR article. The ultimate goal is to find what global topics are covered by the MLR, so those can be used to tag or label all MLR articles. The *Monthly Labor Review* (MLR) is the principal journal of fact, analysis, and research published by the BLS. More information, including published articles, can be found at https://www.bls.gov/opub/mlr/. Articles by economists, statisticians, and other experts from BLS and stakeholders provide a wealth of knowledge on subjects pertaining to a wide range of economic issues. The purpose of this paper is to describe the Shiny apps and not to describe the example data set. Interested readers are referred to Powers, et al., [2021].

## 2. The `gendata` Application

A `growclusters` package user may occasionally wish to generate synthetic data with a known clustering structure to explore the performance of `growclusters` but may not have a properly formatted or appropriate dataset to use as input. When this occurs, it would be helpful to have a tool that generates a dataset to be used as input for the `dpGrowclusters` function. For this reason, we designed the `gendata` Shiny application.

The user opens the `gendata` application to find the `Welcome Tab`. This offers some useful introductory explanations about the options available on the other tabs. The second tab is the `Simple Plot and Table Tab` (see Figure 1). There are several input options that allow the user to customize the dataset that is produced. The user may leave the defaults in place for most of the inputs, however the user is required to specify the number of clusters and the size of the clusters for a dataset to be generated.

The `Number of Clusters` input is the number of clusters the user wishes to be generated in the dataset. The user can specify any numbers of clusters, with a minimum number of two. There is also an input for the `Size of the Clusters`, where the user must input $x$ comma separated values that sum to one. The number of values must equal the specified number of clusters. For example, if the user wants three clusters, an input of (0.4,0.3,0.3) would be a proper input, since there are three values that sum to 1.0, and each entry determines the number of observations in each cluster. For example, one cluster would have 40% of the points, and the other two would each have 30% of the points.

Additionally, there are several other optional user inputs. These include the `Population Size`, which allows the user to control the number of observations $n$ that are produced. The `Number of Dimensions` corresponds to the number of variables (or columns in the typical data tables used in statistics) in the generated dataset. The user may specify the `noise_scale`, which is the global standard deviation, a factor that affects the elements that are generated. The user may specify whether the seed used for generating the dataset is null or non-null. A null seed will produce a random dataset different from an existing one already generated in the application. Whereas a non-null seed will create a dataset reproducing the previous one. This may prove to be helpful if the user wishes to see the effects of changing other individual variable inputs.

Once the user has set the required inputs, they can press the `Generate Data` button, and a dataset is produced. The dataset is captured and held in the R working space or environment (unless manually cleared) for use as input for functions and applications, including those provided in the `growclusters` package.

After the data are generated, the user can click `Select Variables`. This produces a drop-down menu of variable names for the user to select from for visualization purposes. Once the user selects the variables (or columns in the data matrix), they can click `Visualize Data` and the scatterplot matrix is produced. The user can click `Download File` to save the synthetic data file to their computer for future use.

A third tab entitled `Hierarchical Plot and Table` produces a hierarchical version of a synthetic dataset that has known local groups. In other words, the dataset is made up of a collection of (local) data clusters. A collection of multivariate synthetic datasets is generated, where each multivariate observation is assigned to a local cluster in the data set, which, in turn draws its cluster centers or locations from those of a global partition.

Another innovation of the `growclusters` package is the option to include sampling weights when partitioning or clustering the data. Thus, the `gendata` application allows for generating datasets from such complex sample designs. The ability to generate datasets from a stratified sampling or probability proportional to size (PPS) sampling design is available in the Shiny application described in this paper.

## 3. The dpGrowclusters Application

The dpGrowclusters application allows the user to perform the clustering analysis for what we call the single-source clustering viewpoint. For that clustering context, we assume that the data do not have inherent sub-domain structures. If the user does not have a dataset to work with, then they could randomly generate one using the gendata Shiny application, as discussed in the previous section.

The dpGrowclusters application contains five tabs. When the application opens, the tab displayed is the Welcome tab, similar to the gendata application. This tab simply describes the functionality of the other four tabs and gives the user some helpful background information.

The user would next click on the Load Data tab (see Figure 2). The user can choose to load the data from their computer storage or from the R workspace. If they choose the former, they browse their directory and select the RDS file. If they choose the workspace option, then a list of appropriate data objects in the working space is presented in a drop-down menu, and the user selects their desired data object. The variables for the selected dataset appear in the variable selection drop-down box. The user chooses the variables they wish to visualize, then clicks the View Matrix Plot/Display Data button, and the data table and scatterplot matrix plot appear. At this point, no clustering has taken place. The purpose of the matrix plot is to allow the user to examine the data and visually explore the data for groupings.

To cluster the data, the user clicks on the dpCluster tab (see Figure 3). Here, clustering is performed on the data that was loaded in the previous tab. The user chooses one of three methods to determine the optimal clustering parameter, which in turn influences the estimated number of clusters [Savitsky, 2016]. The three methods are: cross-validation, the silhouette statistic, and the Calinski-Harabasz statistic [Martinez, et al., 2011]. The user can also specify other inputs, including the maximum number of clusters allowed and the *K* value for *k*-fold cross-validation. Once the user selects a method and runs the algorithm, a grouping or partition is produced. A bar plot shows the estimated number of clusters found. The bar heights indicate the number of observations in each cluster. When the algorithm is run, the input for the other tabs is also created, behind the scenes.

The user may then switch to the Scatter Plot tab (see Figure 4.) The Scatter Plot tab shows a matrix plot where the colors indicate cluster or group membership. The user can now visually explore the cluster results in this plot (as opposed to the matrix plot in the Load Data tab, which does not yet have the clusters available at that point.) The user can specify what variables to display, just as they did in the Load Data tab. Both this tab and the fifth tab are dependent upon clustering having been performed in the dpCluster tab. If the clustering has not been performed, there is nothing to display.

The fifth tab (see Figure 5) is titled the Parallel Plot tab. This shows the data in a parallel coordinates plot. Each broken line is one observation (or document in the MLR example) in the dataset. Bundles of lines with similar pathways through the vertical coordinate axes indicate good clusters or that the observations are close together. As with the Scatter Plot tab, the colors indicate cluster membership determined from dpCluster. The user can also choose to highlight and view a single cluster by graying out the others (see Figure 6). Note that the colors assigned to cluster IDs are not the same in the scatterplot matrix and the parallel coordinates plot.

## 4. The `hdpGrowclusters` Application

In some cases, we might have data sets that have a known inherent group structure. For example, observations could represent articles published in annual volumes of a journal as was the case with the MLR data set. With the MLR data, the known groups correspond to the year of publication. The original paper by Savitsky [2016] discusses another motivating application, where the known sub-domain structures are the NAICS codes assigned to establishments.

Using the MLR articles to illustrate the concepts in the following discussion, we could ignore the known sub-domains (or year of publication) and apply clustering to the articles as if they were published at the same time. This is the single-source clustering approach described in the previous section.

If we use the single-source clustering approach, then we are not accounting for the fact that articles were published over a period of years. One option to account for the year of publication would be to separately apply the single-source clustering to the articles published in each individual year. However, global topics would be difficult to determine by merging 'local' topics and to label them across the years. So, this is not appropriate for the goal of finding global topics we could use to label all MLR articles with their topics, regardless of the year.

We would like to find global clusters for all MLR articles in the corpus *and* account for possible dependencies based on the year of publication. This is the purpose behind the original hierarchical clustering approach by Savitsky [2016]. Global clusters that account for possible local dependencies among known sub-domains are called *hierarchical* `growclusters`. This should not be confused with *hierarchical clustering*, which is a well-known clustering method that has been around since at least the 1960s; see Martinez, et al. [2011]. The hierarchical `growclusters` method is inspired by Bayesian hierarchical models but is not Bayesian.

We developed a Shiny application called `hdpGrowclusters` similar to `dpGrowclusters`. There are analogous ways to obtain optimal parameters (cross-validation, Calinski-Harabasz, silhouette), which will appear in the hierarchical version. The additional information on sub-domains gives rise to different types of visualizations to be included in the `hdpGrowclusters` app. The first five tabs work very similarly to the five tabs described in the previous section. One additional feature of this app is the ability to choose matrix plots of individual domains in addition to the entire dataset, for both the `Load Data` tab and the `Scatter Plot` tabs (see Figures 7 and 8).

A sixth tab, called `Conditional Barplots,` is new and unique to the `hdpGrowclusters` app. It also exhibits the additional functionality that is available by adding the domain variable. The user can choose to either display the distribution of cluster IDs in each domain (see Figure 9) OR display the distribution of domains in each cluster (see Figure 10).

Additionally, a reset feature has been added to each of the three apps. That is, when the reset occurs, the previous work is erased, and the user essentially starts from scratch. This is necessary when the user wants to create multiple datasets in the `gendata` app, or to analyze multiple datasets in the other two apps. This enables the user to continue without having to shut down and restart the app. Since an independent dataset is created in each of the final four tabs in the `gendata` app, it makes sense to have a separate reset button for each tab. Since the final three tabs of the `dpGrowclusters` app and final four tabs of the `hdpGrowclusters` app are all dependent on the initial data loaded all work on the one loaded dataset, it made sense to have the reset feature be tied to the `Load Data` tab. That is, when the user chooses a different data load method, or chooses to load a different dataset, the previous results are removed, and the user starts with a blank slate.

## 5. Future Work and Final Comments

Our next step is to finalize the programming, including error checking, on all three apps. We will then write several vignettes illustrating the functionality of the `growclusters` package and its accompanying R Shiny apps. We plan to submit the package to CRAN and/or GitHub once it is ready for public release.

This paper focuses on the development of R Shiny apps and not so much on the functionality included in the `growclusters` package. The approach by Savitsky [2016] allows for clustering of data collected using a complex sample design and incorporates sampling weights. Future work will explore this additional clustering framework with additional data sets and will illustrate its properties.

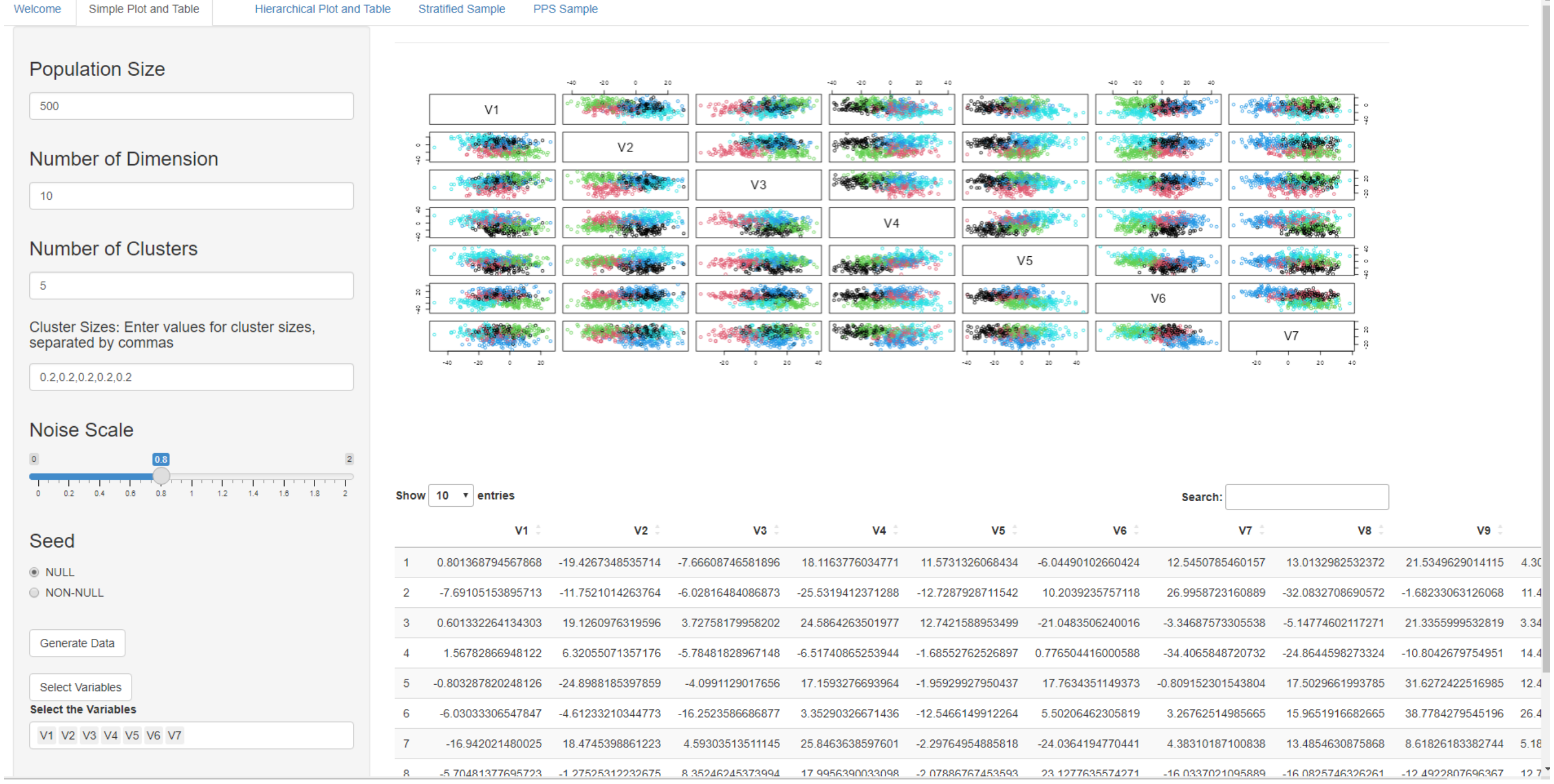

**Figure 1**. This shows the `gendata` application `Simple Plot and Table` tab. The user sets the inputs, generates the data, chooses the variables, and then generates the matrix plot.

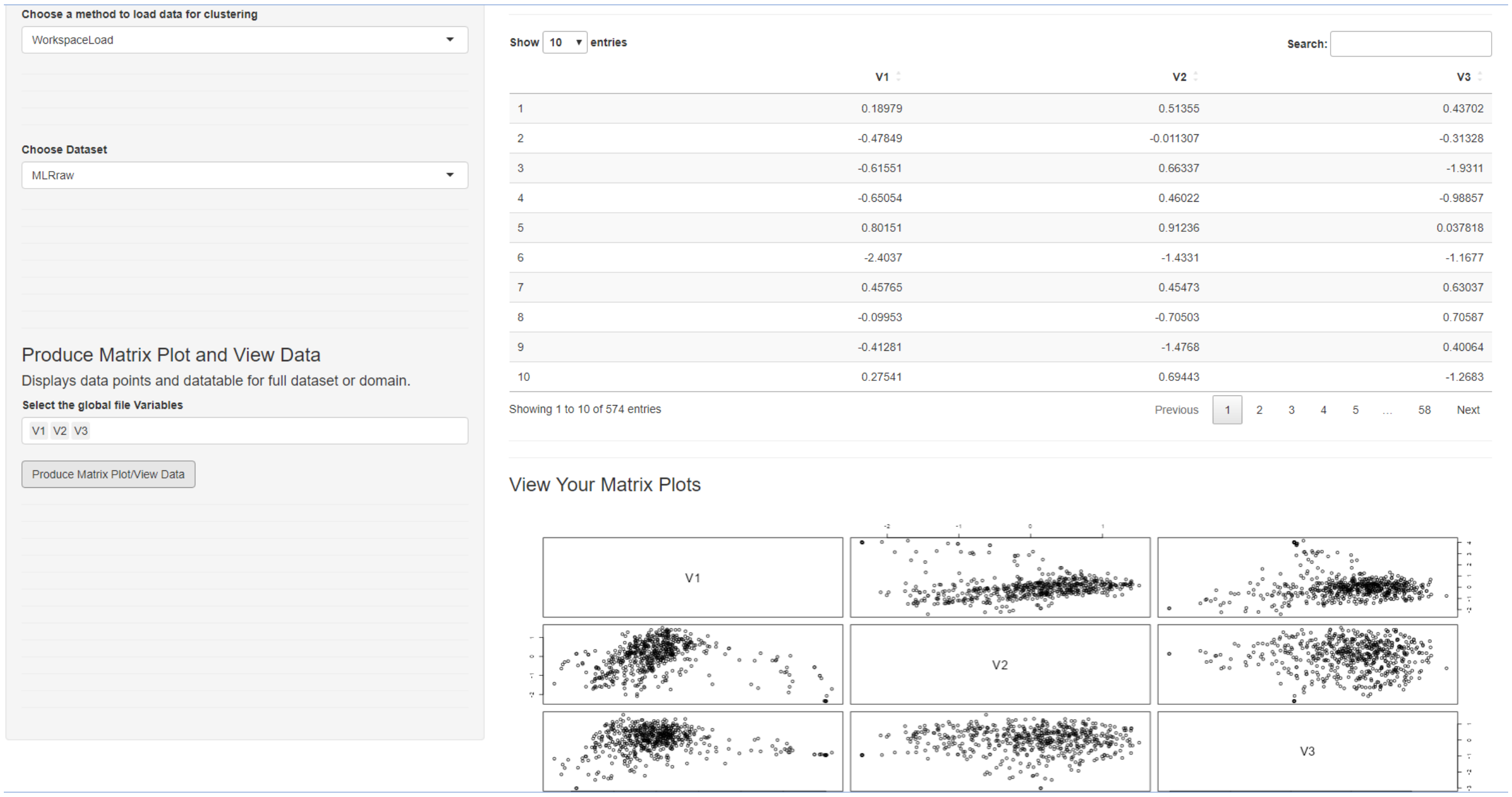

**Figure 2**. This is a screenshot of the `dpGrowclusters` application `Load Data` tab. Note that the user has the option to load a data set stored in a computer directory or one that is already loaded into the workspace.

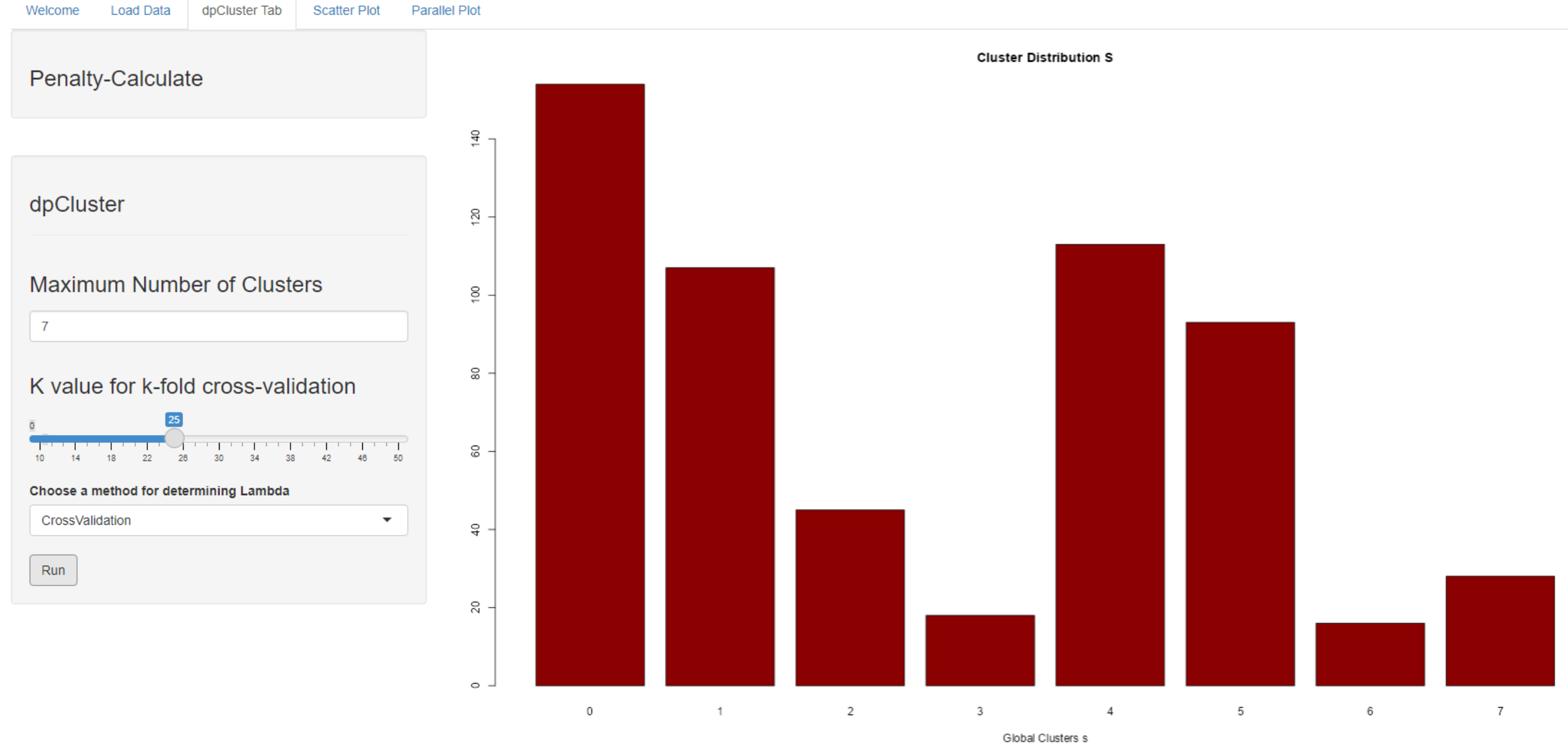

**Figure 3**. Here is a screenshot of the `dpCluster` tab, showing a bar plot distribution of the clusters using the MLR data. The user can choose the method for determining the optimal lambda parameter on the left panel. The parameter value impacts the estimated number of clusters.

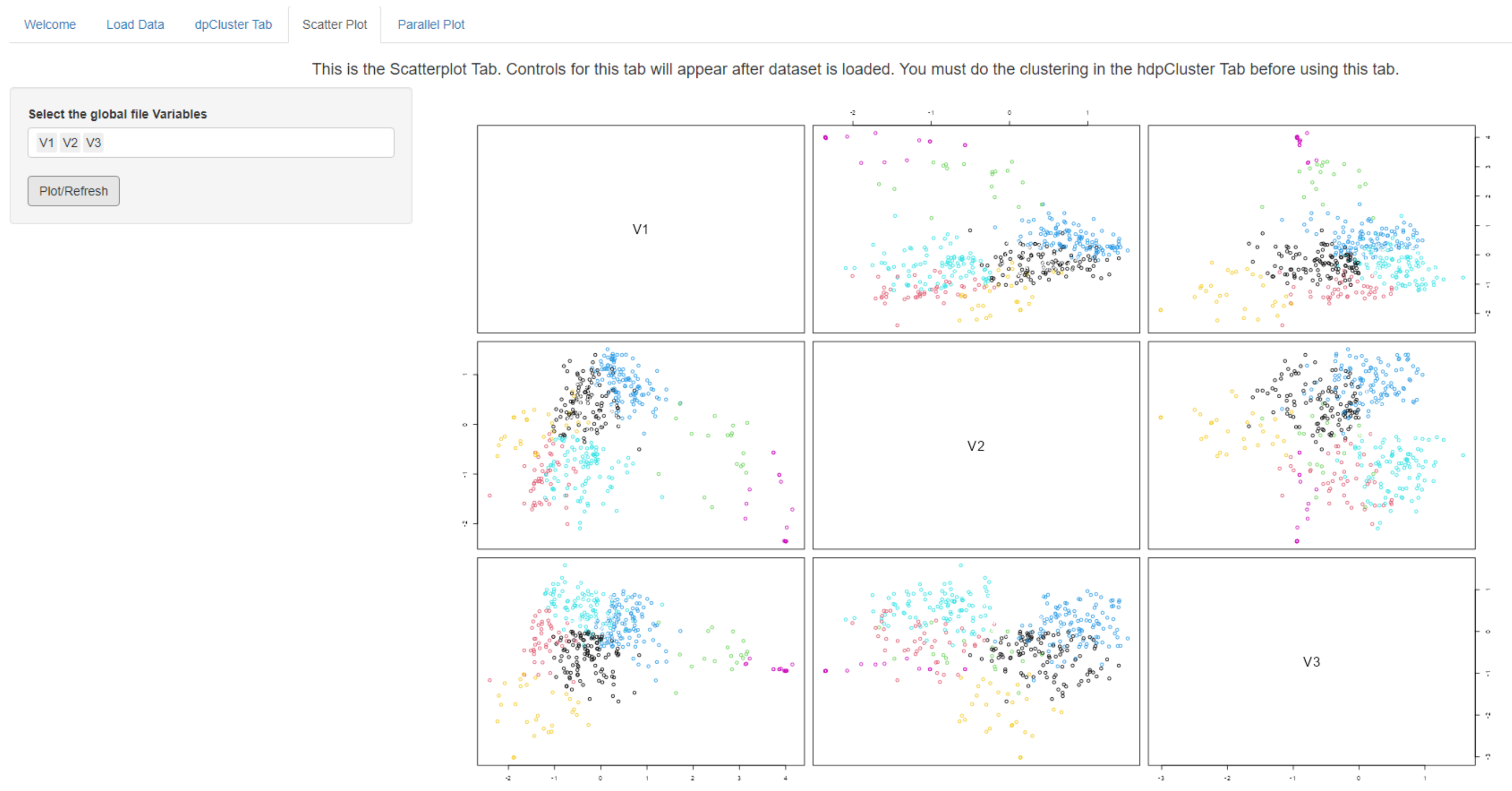

**Figure 4.** The cluster results can be viewed in the `Scatter Plot` tab. Here we see clusters found in the MLR data. The colors indicate the cluster, and the groups seem visually reasonable.

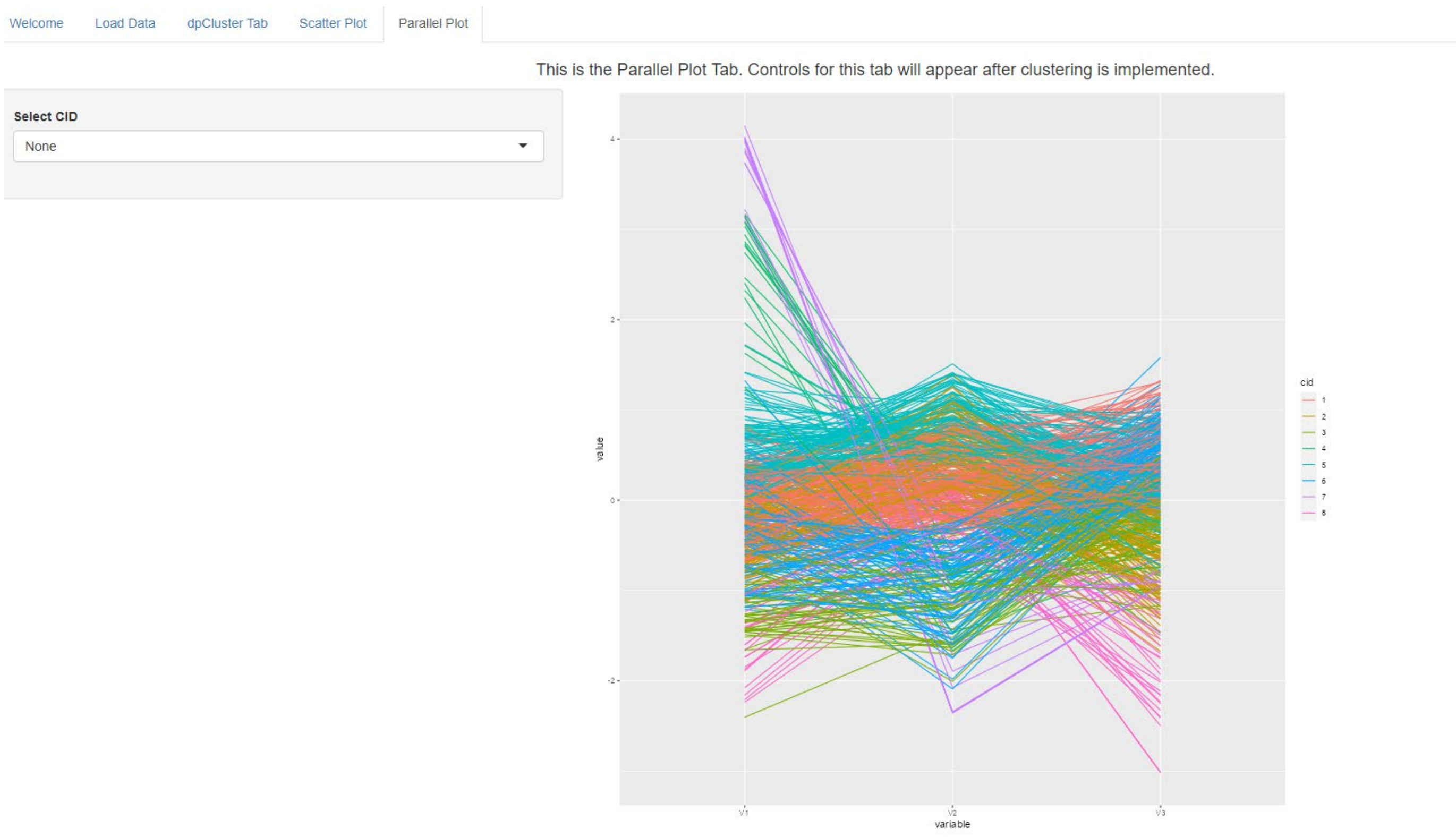

**Figure 5.** The clusters can be viewed in parallel coordinates at the `Parallel Plot` Tab. This shows the clusters found in the MLR (raw) data, where a color indicates the cluster.

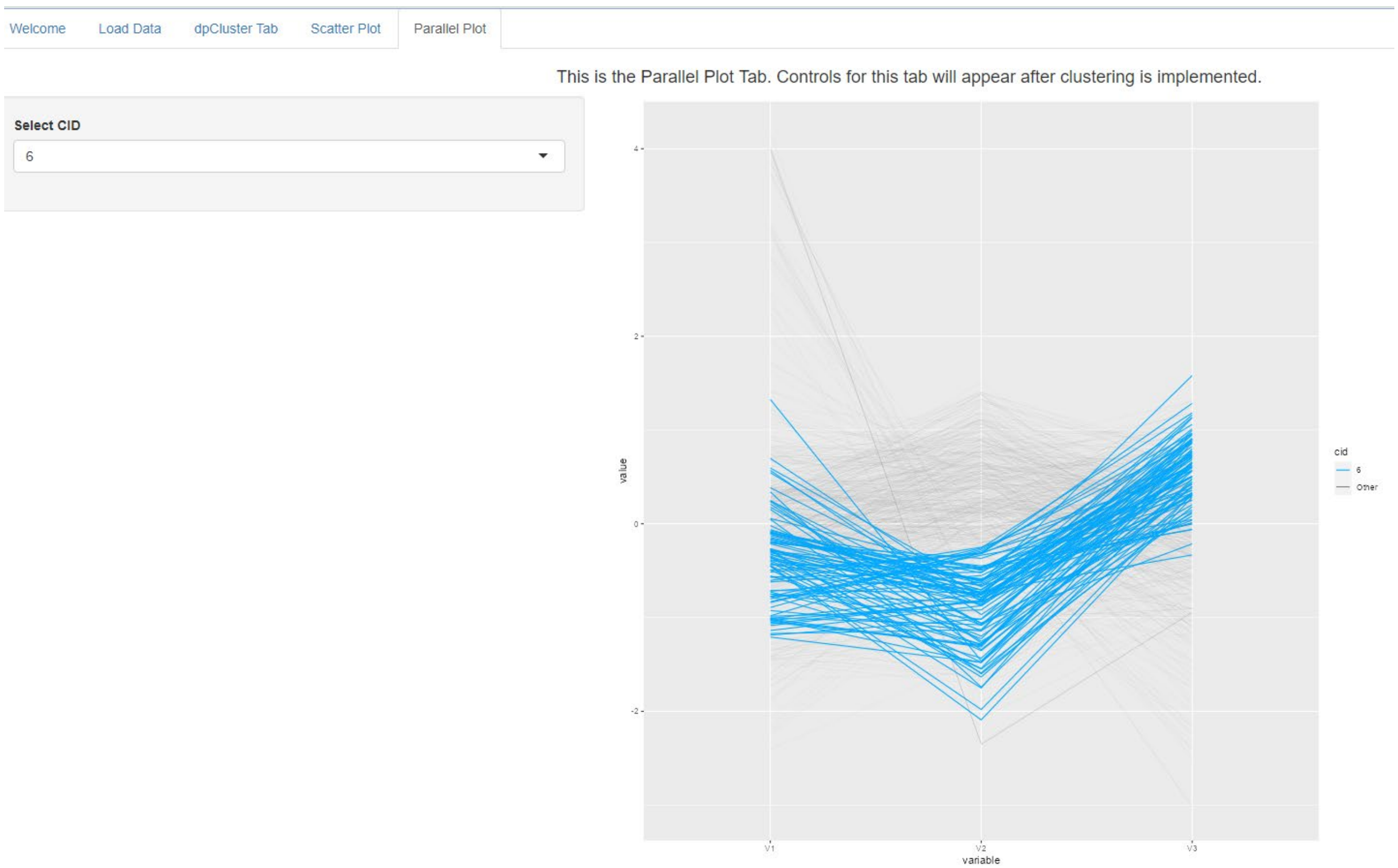

**Figure 6**. In the `Parallel Plot` Tab, the user can select a cluster to visualize in the dropdown box at the left. This allows one to see if the broken lines in a cluster follow a similar path through the axes and/or are bundled together, indicating observations in the cluster are similar to each other.

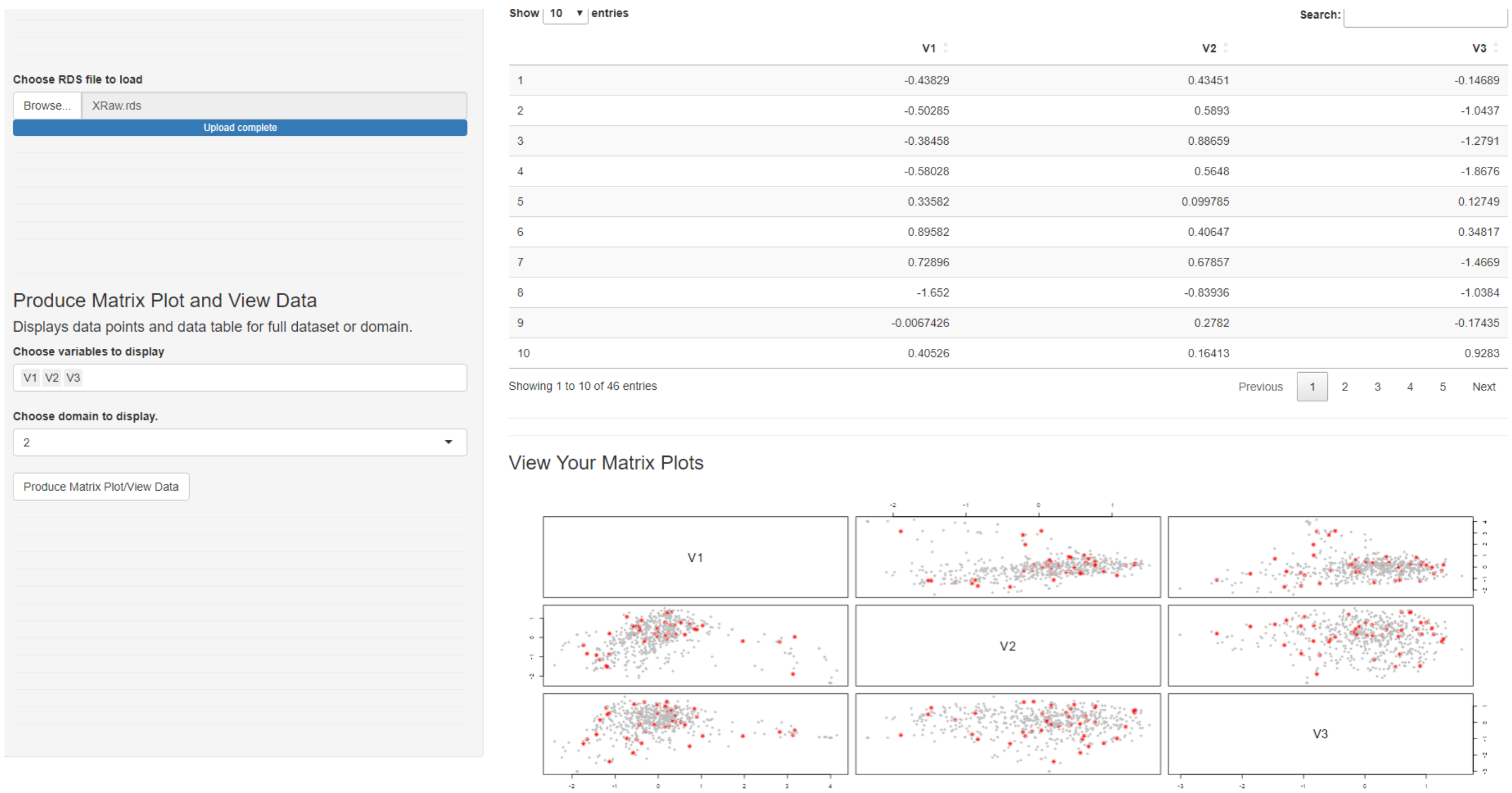

**Figure 7**. This is a screenshot of the `hdpGrowclusters` application `Load Data` tab. Note that though the results of this tab are simliar to those os the corresponding tab in the `dpGrowclusters` application `Load Data` tab (Figure 2), this one has the additional feature of individual domain selection. Here Domain 2 has been selected, and all other values are from other domains are grayed out.

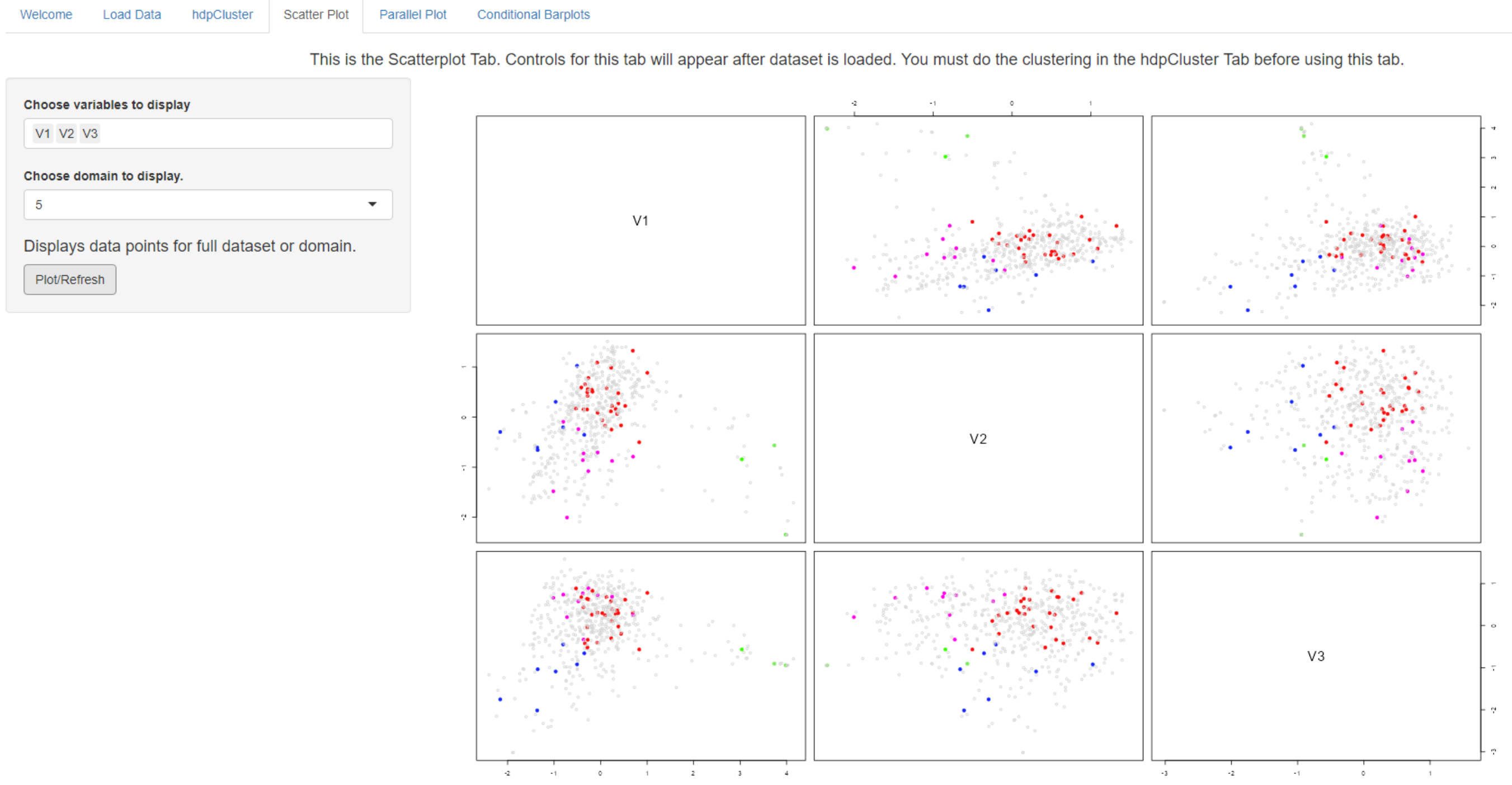

**Figure 8**. This is a screenshot of the `hdpGrowclusters` application `Scatter Plot` tab. Note that though the results of this tab are simliar to those of the corresponding tab in the `dpGrowclusters` application `Scatter Plot` tab (Figure 4), it has the additional feature of individual domain selection. Here Domain 5 has been selected, and all other values are from other domains are grayed out. This matrix plot is also color coded, with each color representing a cluster.

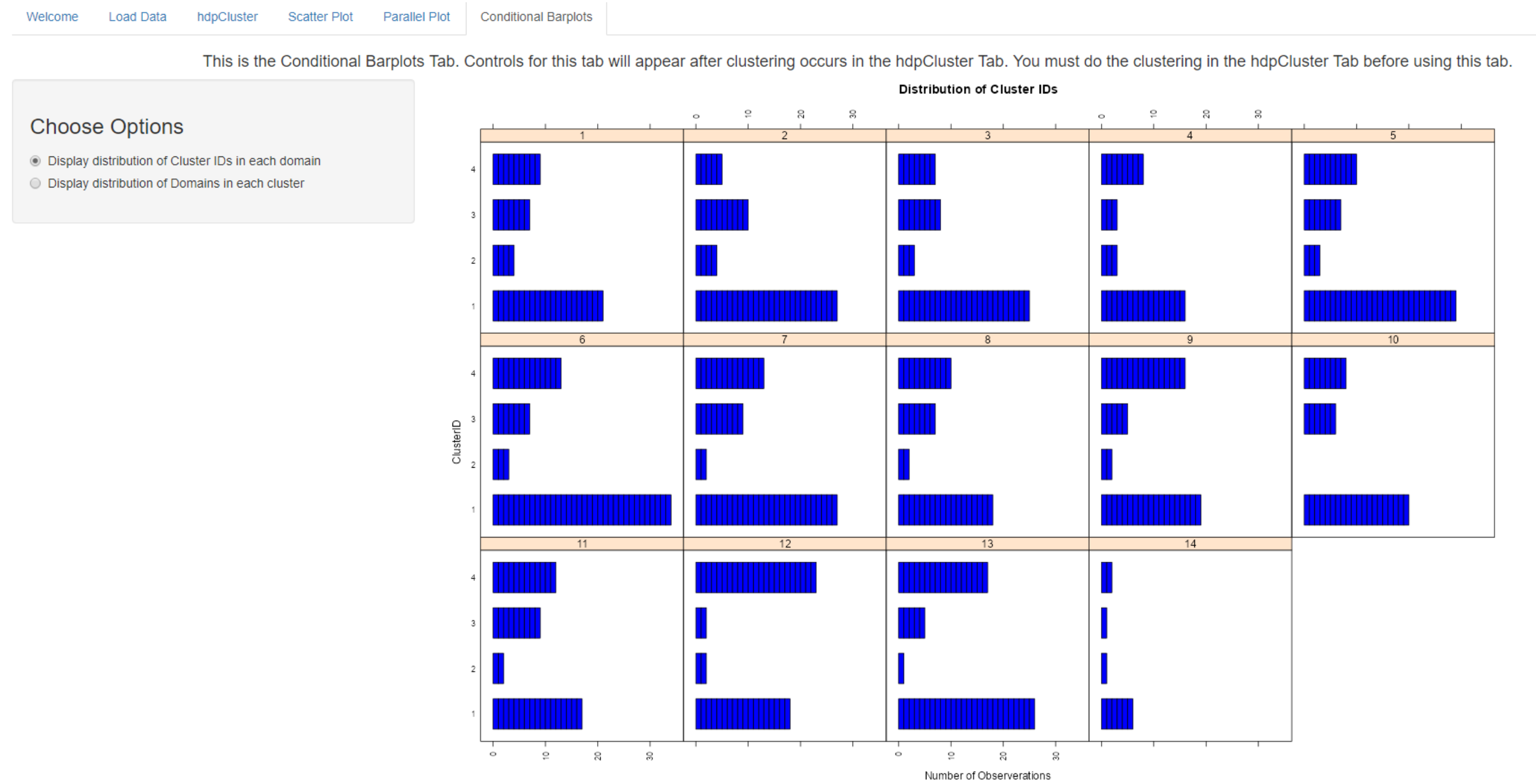

**Figure 9**. This is a screenshot of the `hdpGrowclusters` application `Conditional Barplot` tab. When the first button is selected, we see the distribution of cluster id's in each domain.

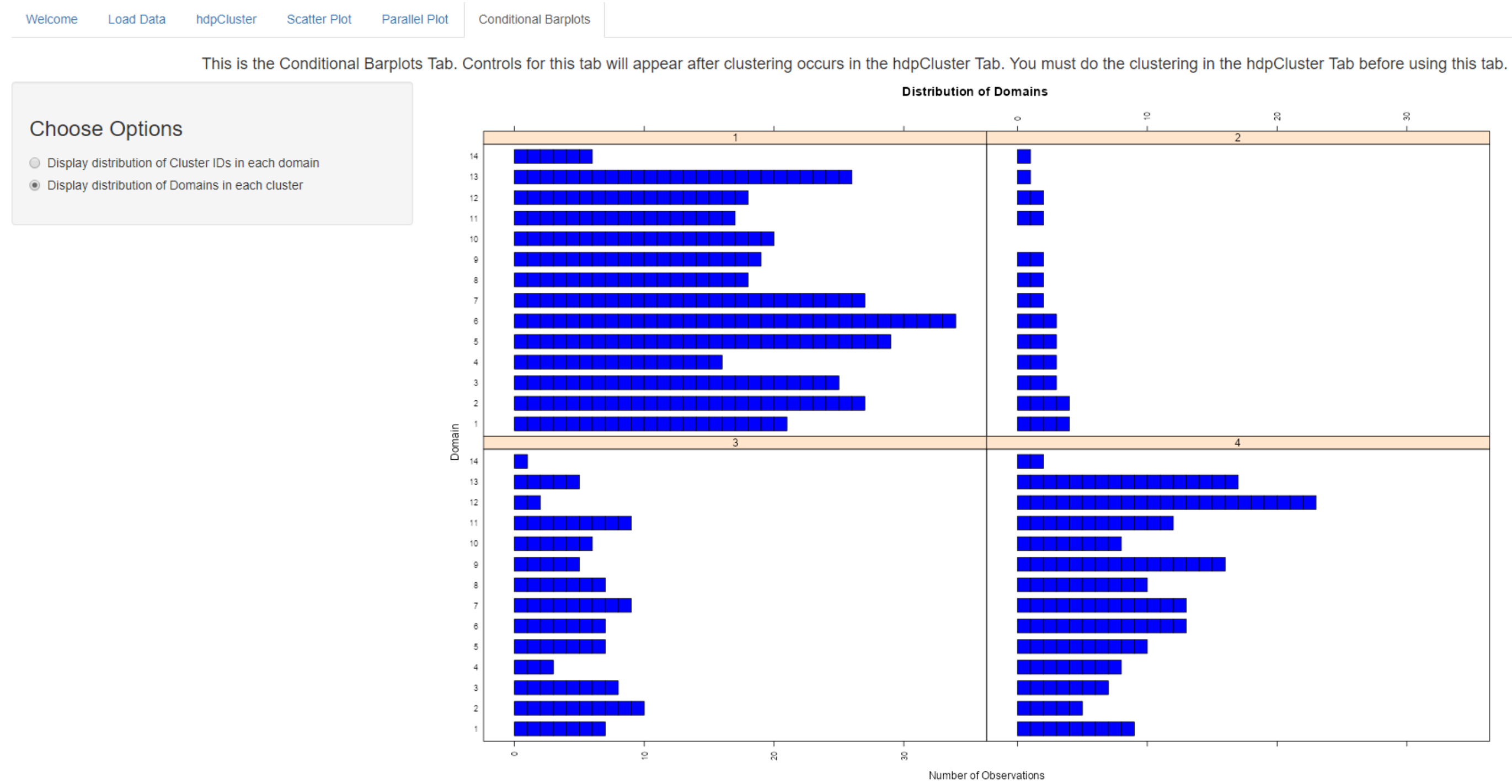

**Figure 10**. This is a screenshot of the `hdpGrowclusters` application `Conditional Barplot` tab. When the second button is selected, we see the distribution of domains in each cluster.